\documentclass[a4paper,american,journal=ancac3, manuscript=article]{achemso}
\usepackage[T1]{fontenc}
\usepackage[latin9]{inputenc}
\usepackage{amssymb}
\usepackage{graphicx}
\usepackage{esint}

\makeatletter

\title{Plasmonic Nano-Gap Tilings: Light-Concentrating Surfaces for Low-Loss
Photonic Integration}
\author{Paul M. Z. Davies, Joachim M. Hamm, Yannick Sonnefraud, Stefan A.
Maier and Ortwin Hess}
\affiliation{The Blackett Laboratory, Department of Physics, Imperial College
London,\\South Kensington Campus, London SW7 2AZ, UK}
\email{o.hess@imperial.ac.uk}

\makeatother

\usepackage{babel}
\begin{document}
\begin{abstract}
Owing to their ability to concentrate light on nanometer scales, plasmonic
surface structures are ideally suited for on-chip functionalization
with nonlinear or gain materials. However, achieving a high effective
quantum yield across a surface not only requires strong light localization
but also control over losses. Here, we report on a particular class
of tunable low-loss metasurfaces featuring dense arrangements of nanometer
sized focal points on a photonic chip with an underlying waveguide
channel. Guided within the plane, the photonic wave evanescently couples
to the nano-gaps, concentrating light in a lattice of hot-spots. In
studying the energy transfer between photonic and plasmonic channels
of single trimer molecules and triangular nano-gap tilings in dependence
on element size, we identify different regimes of operation. We show
that the product of field enhancement, propagation length and element
size is close-to-constant in both the radiative and subwavelength
regimes, opening pathways for device designs that combine high field
enhancements with large propagation lengths.\end{abstract}
\maketitle

\begin{list}{KEYWORDS:}
\item Nano-gap tiling, Low-loss plasmonics, Silicon photonics, Waveguide integration, Mode hybridization 
\end{list}

\begin{figure}[H]
\begin{centering}
\includegraphics[width=0.75\textwidth]{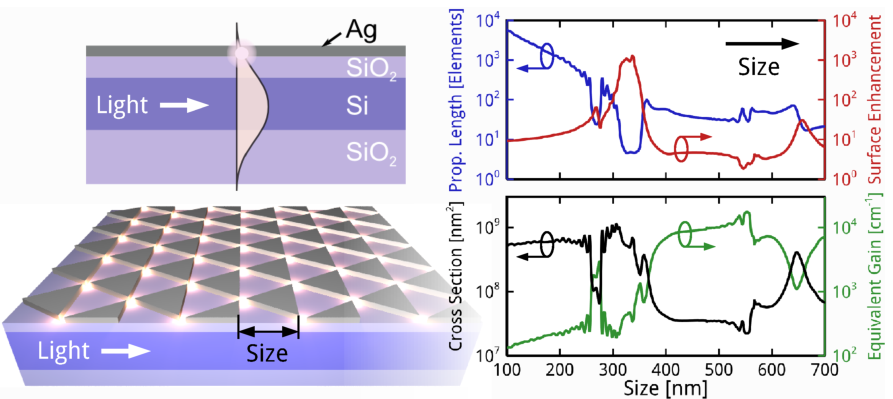}
\par\end{centering}
\end{figure}

Mediated by subwavelength interactions with metallic particles, nanoplasmonic
surfaces can alter the localization and propagation characteristics
of light at interfaces.\cite{Murray2007,Radko2009,Schuller2010} The
steady progress in lithographic\cite{Henzie2009,Boltasseva2009} and
self-assembly techniques\cite{Stebe2009,Fan2010} makes it possible
to shape metallic surface elements with increasing precision, reaching
towards sub-nm scales.\cite{Duan2012} In the wake of this technological
development, schemes involving coupled plasmonic particle clusters
and arrays have been designed, fabricated and studied, exploring their
potential to slow or stop light, \cite{Tsakmakidis2007,Zhang2008,Liu2009a}
to nano-focus light in the gaps between or at the edges of the nano-particles,\cite{Kravets2010,Zhou2011}
to resonantly transport energy along particle chains,\cite{Maier2003b}
to control the phase,\cite{Yu2011a,Yu2012,Ni2012a} polarization\cite{Yu2012,Lin2013}
and to refract light at negative angles.\cite{Yu2011a,Ni2012a}

As research continues to explore nanoplasmonic surfaces with engineered
optical properties and their functionalization with nonlinear and
quantum gain \cite{Hess2013,Hess2012} materials as diverse as gases
\cite{Kim2008}, semiconductors\cite{Tanaka2010}, dyes \cite{Xiao2010}
and graphene\cite{Kravets2013} there is a growing gap between the
wealth of potential applications and practical concepts for on-chip
integration with photonic technologies. In fact, until now, plasmonic
and metamaterial surfaces are mostly operated out-of-plane; \textit{i.e.}
by illumination under normal incidence or evanescent excitation with
near-field microscope tips;\cite{Hecht1996,Krenn2002,Brun2003} two
well-established techniques that are convenient to perform, yet with
limited potential for on-chip integration. Fundamentally, integration
with photonic circuitry requires keeping light confined and in interaction
with the structured surface, without inducing uncontrolled scattering
or excessive dissipative loss. The usual trade-off is that a stronger
confinement of light to the metallic surface increases attenuation
and heat generation due to Ohmic losses.\cite{Takahara1997,Bozhevolnyi2005,Oulton2008a}
Hybrid metal-dielectric waveguide structures \cite{Holmgaard2007,Oulton2008,Ditlbacher2008,Sorger2011}
offer one possibility to break the interdependency of loss and localization,
but require careful design to achieve good loss-performance and compatibility
with established photonic architectures. Furthermore, from a fabrication
and integration perspective, it is highly desirable to establish a
platform where light waves, guided within a photonic layer, can be
manipulated on subwavelength scales using metallic nano-structures
on the surface.\cite{Sidiropoulos2012} Such hybrid platforms may
not only provide well-defined interfaces for on-chip integration of
plasmonic devices (\textit{e.g}. sensors or spectroscopic elements)
with photonic signal-processing circuits, but also offer practical
routes towards active nanoplasmonic components that exploit the enhanced
surface-interaction with fluorescent,\cite{Fort2008} magneto-optic\cite{Jain2009}
or nonlinear materials.\cite{Kim2008}

\begin{figure}[H]
\begin{centering}
\includegraphics[width=0.75\textwidth]{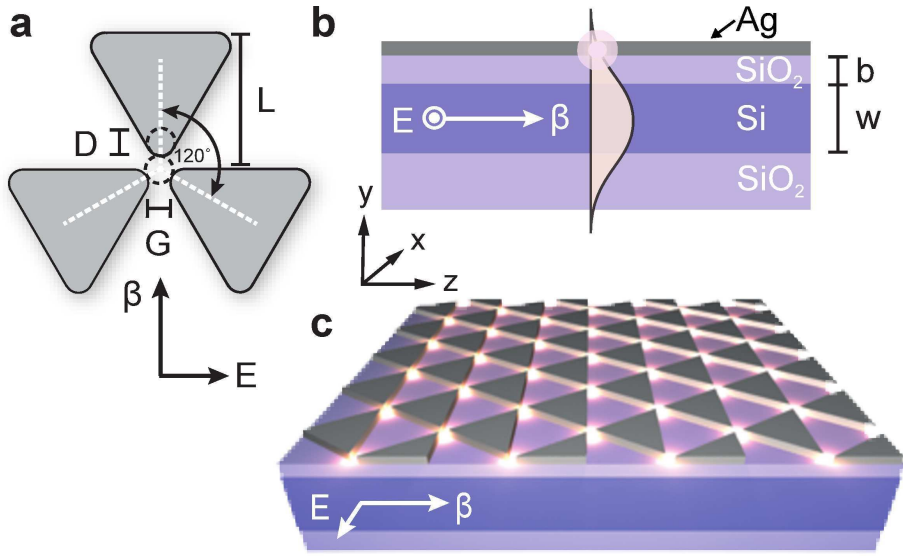}
\par\end{centering}
\caption{\label{fig:1-Structure}Triangular nano-gap tiling made of $C_{3v}$
symmetric trimer molecules (a) that are densely arranged on the top
of a dielectric stack (b). The TE-polarized photonic field propagates
with $\beta=n_{eff}k_{0}$ inside the silicon waveguide ($w$) while
being evanescently focused into bright spots on the surface (c). Both
gap and corner diameter of the elements are set to scale with the
pitch $L$ (\textit{i.e.,} $G=D=L/10$).}
\end{figure}

In this article we report on a particular class of integrated plasmonic
surface structures, plasmonic nano-gap tilings (NGTs), featuring dense
lattices of focal points (\ref{fig:1-Structure}). To understand the
general properties of plasmonic NGTs, we theoretically study one of
the simplest structures: triangular surface patches of fixed orientation
arranged in a triangular lattice. The chosen combination of shape
and lattice was guided by our aim to concentrate light in a dense
array of hot-spots to achieve a strong enhancement of the surface
fields. 

The surface structure is fully characterized by the pitch $L=a\sqrt{3}/2$,
given by the projection of lattice vectors $\vec{a}_{1}$ and $\vec{a}_{2}$
in the propagation direction, the gap $G$ and the element corner
rounding of diameter $D$ (\ref{fig:1-Structure}a). The triangular
NGTs borrow symmetry and light-enhancing properties from the trimer
molecule. As shown in \ref{fig:1-Structure}(a) both the $C_{3v}$
symmetry of the molecule and two sets of directional axes of excitation
($+z$ and $-z$) are retained in the assembled 2D tiling (\ref{fig:1-Structure}c).

We consider $30\,\mathrm{nm}$ thick silver ($\mathrm{Ag}$) structures
on top of a planar dielectric stack consisting of silicon oxide ($\mathrm{SiO}_{2}$)
substrate, a planar $w=220\,\mathrm{nm}$ thick silicon ($\mathrm{Si}$)
waveguide, and a $\mathrm{SiO}_{2}$ buffer layer of tunable thickness
$b$ (\ref{fig:1-Structure}b). To maintain the lateral mode profile,
we operate at a fixed wavelength of $\lambda_{0}=1550\,\mathrm{nm}$
where the guiding $\mathrm{Si}$ layer supports only the fundamental
transverse electric (TE) and magnetic (TM) modes. At this wavelength
the complex permittivity of the silver\cite{Johnson1972} can be approximated
by $\varepsilon_{\mathrm{Ag}}=-127-3.38i$.

\begin{figure}[H]
\begin{centering}
\includegraphics[width=0.75\textwidth]{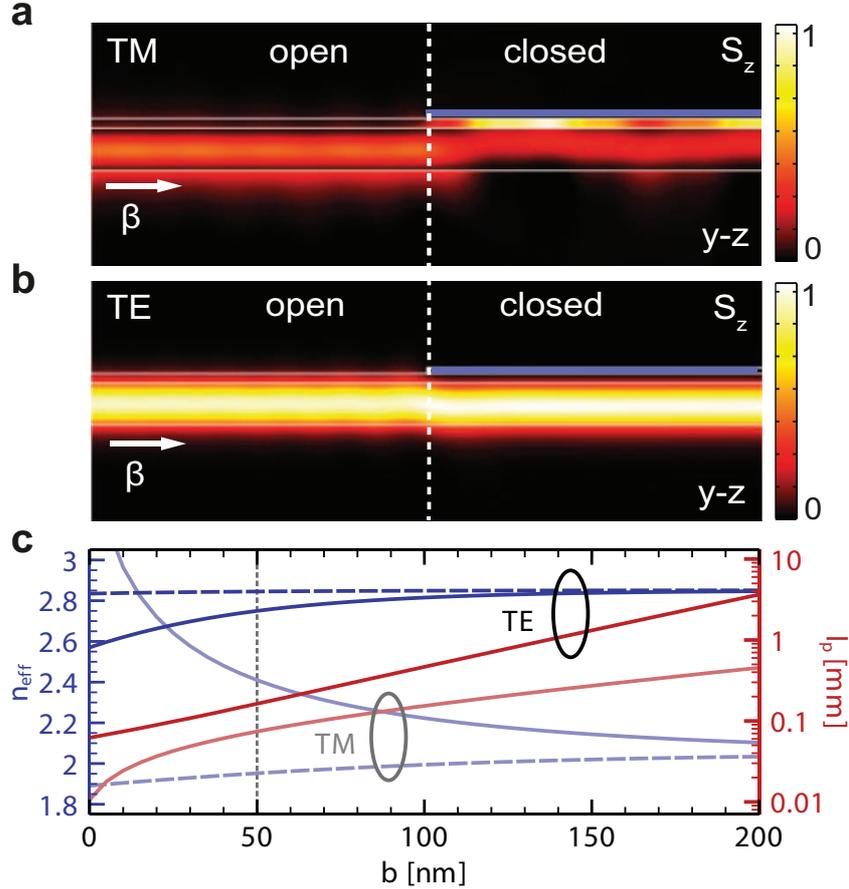}
\par\end{centering}

\caption{\label{fig:2-Open/Closed}(a)+(b) $z$-component of the Poynting vector
for TM (a) and TE (b) modes scattering at the edge of a $30\mathrm{nm}$
thick silver layer separated by a $\mathrm{SiO_{2}}$ buffer of height
$b=50\,\mathrm{nm}$ from the $\mathrm{Si}$ waveguide. (c) In directly
exciting a plasmonic mode in the buffer, the TM mode experiences a
strong mismatch of effective indices (light blue lines) and a consistently
lower propagation lengths (light red lines) than the TE mode (dark
lines) for equivalent buffer sizes $b$. Dashed lines correspond to
the open, solid lines to the closed waveguide sections.}
\end{figure}

\subsubsection*{RESULTS AND DISCUSSION}

\paragraph*{Waveguide Integration.}

To characterize the fundamental properties of the hybrid platform,
we first study modes impinging on the interface between the open (\textit{i.e.}
uncovered) and closed (\textit{i.e.} metal covered) waveguide sections.
For TM polarization (\ref{fig:2-Open/Closed}a), the mode hybridizes\cite{Ditlbacher2008}
with the surface plasmon polariton (SPP) mode at the metal-buffer
interface, concentrating the energy in the low index buffer.\cite{Oulton2009,Dai2009,Alam2010}
As a consequence, both the mode profiles and the effective refractive
index (\ref{fig:2-Open/Closed}c; light blue lines) between the two
sections mismatch strongly; in particular for $b<100\,\mathrm{nm}$.
In contrast, the TE mode (\ref{fig:2-Open/Closed}b) retains its photonic
character across the discontinuity and, while being slightly pushed
down into the low-index substrate, experiences a considerably smaller
perturbation of its profile and a change in its effective index (\ref{fig:2-Open/Closed}c;
dark blue lines). More specifically, at $b=50\,\mathrm{nm}$, the
effective refractive index mismatch, $\Delta n_{\mathrm{eff}}$, between
the open and closed sections for a TE excitation amounts to $-0.1$,
compared to $+0.45$ in the TM case. We also note that the propagation
length $l_{p}^{\mathrm{TE}}$ of the TE mode is consistently larger
than that of the TM mode and rises strictly exponential with increasing
buffer height (\ref{fig:2-Open/Closed}c; red lines, right axis).
This exponential behavior at small buffer heights indicates that,
for TE excitation, losses are induced by the evanescent tail of the
mode profile touching the metallic surface rather than the direct
coupling in the case of the TM mode. At $b=50\,\mathrm{nm}$, we record
a value of $l_{p}^{\mathrm{TE}}\approx134\,\mathrm{\mu m}$; roughly
twice as high as the corresponding TM value confirming that the TE
mode not only features the better matched mode profile, but is also
more favorable in terms of dissipative loss.

\begin{figure}[H]
\begin{centering}
\includegraphics[width=0.75\textwidth]{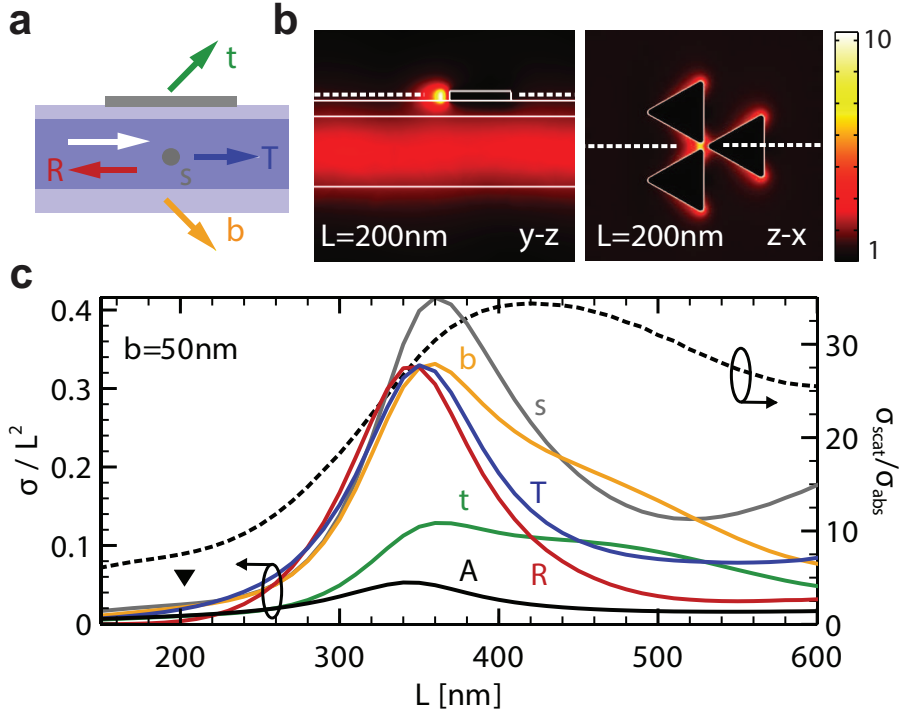}
\par\end{centering}

\caption{\label{fig:3-TrimerScattering}(a) Schematic of the backwards (R;
red), forwards (T; blue), top (t; green), bottom (b; orange) and side-way
(s; gray line) scattering channels of a single trimer antenna; (b)
Electric field amplitude in $y-z$ plane and $z-x$ plane at positions
indicated by the dashed white lines for an isolated trimer nano-antenna
of dimension $L=200\,\mathrm{nm}$; (c) Scattering cross-sections
for the various scattering channels (left axis), absorption cross-section
(black line, left axis) and ratio between scattering/absorption cross-sections
(dashed line, right axis).}
\end{figure}

\paragraph*{Trimer Surface Antenna: Scattering Spectra.}

Field concentration in the focal points is achieved \textit{via} evanescent
interaction of the guided TE mode with surface elements. To explore
this mechanism we first consider an isolated trimer molecule\cite{Koh2011}
(\ref{fig:1-Structure}a) of individual element size $L$. The isolated
trimer can be understood as a plasmonic surface antenna that is evanescently
driven by the photonic field, absorbing and scattering energy into
the various in- and out-of-plane channels (\ref{fig:3-TrimerScattering}a).
\ref{fig:3-TrimerScattering}c shows the scattering and absorption
cross-sections associated with the different channels. The ratio of
scattered energy to the absorbed energy (dashed line, right axis)
shows that the absorption is $5-35$ times lower than scattering,
as the antenna cannot be considered small compared to the effective
wavelength of $\lambda_{eff}\approx544\,\mathrm{nm}$. For $L<200\,\mathrm{nm}$,
in the subwavelength regime, the scattering cross-sections are smaller
than $0.04L^{2}$. With increasing $L$, the various scattering cross-sections
(left axis) rise and peak at $L\approx350\,\mathrm{nm}$, due to the
excitation of localized surface plasmon resonances (LSPRs). Owing
to the strong refractive index mismatch between the air cladding and
silicon waveguide, the top scattering is weakest while the scattering
to the side achieves the highest cross-section of just above $0.4L^{2}$.
Beyond the dipolar resonance $L>400\,\mathrm{nm}$ the scattering
cross-sections drop, except those related to the out-of-plane scattering
(channels $t$ and $b$) that exhibit a characteristic shoulder. This
suggests the presence of another radiative resonance at around $L=500\,\mathrm{nm}$.
The exceptional stability of the driving photonic field, even at resonance,
is one of the characteristics of TE excitation that makes this configuration
inherently suitable for the optically driven chip-integrated plasmonic
antennas.

\begin{figure}[H]
\begin{centering}
\includegraphics[width=0.75\textwidth]{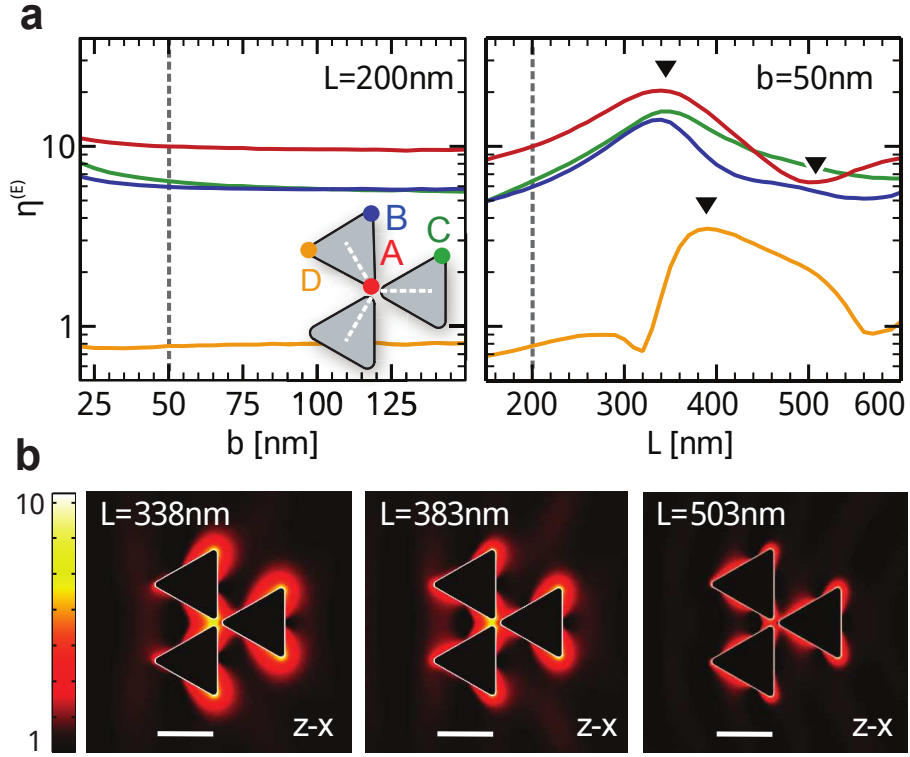}
\par\end{centering}

\caption{\label{fig:4-TrimerEnhance} (a) Surface field enhancement $\eta^{(E)}$
at the four corner points A (red), B (blue), C (green), and D (orange)
in dependence on buffer height $b$ (for fixed $L=200\,\mathrm{nm}$)
and size $L$ (for fixed buffer height $b=50\,\mathrm{nm}$). (b)
Color plots of $\log(\eta^{(E)})$ in $z-x$ plane for $L=338\,\mathrm{nm}$,
$L=383\,\mathrm{nm}$ and $L=503\,\mathrm{nm}$ show the distribution
of intensities.}
\end{figure}

\paragraph*{Trimer Surface Antenna: Field Enhancement.}

Despite the weaker evanescent surface interaction, the TE-polarized
photonic field can lead to strong lateral charge separation within
the metallic surface elements. This can lead to capacitive coupling
in the nano-gaps between the elements inducing strong fields which
peak at resonance.\cite{Muhlschlegel2005,Ghenuche2008}As shown in
\ref{fig:3-TrimerScattering}b, the field concentrates on the surface
into bright spots at the corners of the antenna, while the propagating
field in the waveguide below remains mostly unperturbed.

To quantify the achieved light concentration, we measure the relative
surface enhancement, $\eta^{(E)}$, of the electric field at the different
corners of the trimer (\ref{fig:4-TrimerEnhance}a; inset) by dividing
the field amplitude averaged over a small volume around the respective
corner by the same surface field without the antenna, \textit{i.e.}
$\eta_{\circ}^{(E)}=|E(\mathbf{r}_{\circ})|/|E_{0}|$ ($\circ=A,\, B,\, C,\, D$).
The results in \ref{fig:4-TrimerEnhance} give a strong indication
that the spot-enhancements are largely independent of buffer height
(\ref{fig:4-TrimerEnhance}a; left panel), as the evanescent surface
field is effectively funneled into the central focus point $A$ without
a significant perturbation of the mode propagating within the waveguide.
With increasing size $L$, the antenna (\ref{fig:4-TrimerEnhance}a;
right panel) becomes resonant (at $L\approx338\,\mathrm{nm}$) and
the field enhancement at the center position increases from $\eta_{A}^{(E)}\approx10$
to $\eta_{A}^{(E)}\approx20$. The position of the enhancement peaks
(and dips) depends critically on the position of evaluation. The most
distinct features are the peaks of $\eta_{A}^{(E)}$, $\eta_{B}^{(E)}$
and $\eta_{C}^{(E)}$ at around $L\approx338\,\mathrm{nm}$, the broad
peak of $\eta_{D}^{(E)}$ at $L\approx383\,\mathrm{nm}$, and the
dip of $\eta_{A}^{(E)}$ at around $L\approx503\,\mathrm{nm}$ (see
markers in \ref{fig:4-TrimerEnhance}a). \ref{fig:4-TrimerEnhance}b
shows the two resonant states of the antenna, characterized by a strong
lateral charge separation ($L\approx338\,\mathrm{nm}$; left panel)
and an additional longitudinal charge separation ($L\approx383\,\mathrm{nm}$;
middle panel). With increasing size $L$ additional nodal planes in
the longitudinal direction become visible ($L\approx503\,\mathrm{nm}$;
right panel) as a signature of wave retardation. As the element size
becomes comparable to the effective wavelength of the waveguide mode,
the polarizing electric field changes orientation across the antenna,
exciting higher order (subradiant) modes. This retardation effect
also illuminates the otherwise dark corner $D$ and causes, in weakening
the lateral resonance, a decrease of the enhancement in the center. 

\begin{figure}[H]
\begin{centering}
\includegraphics[width=0.75\textwidth]{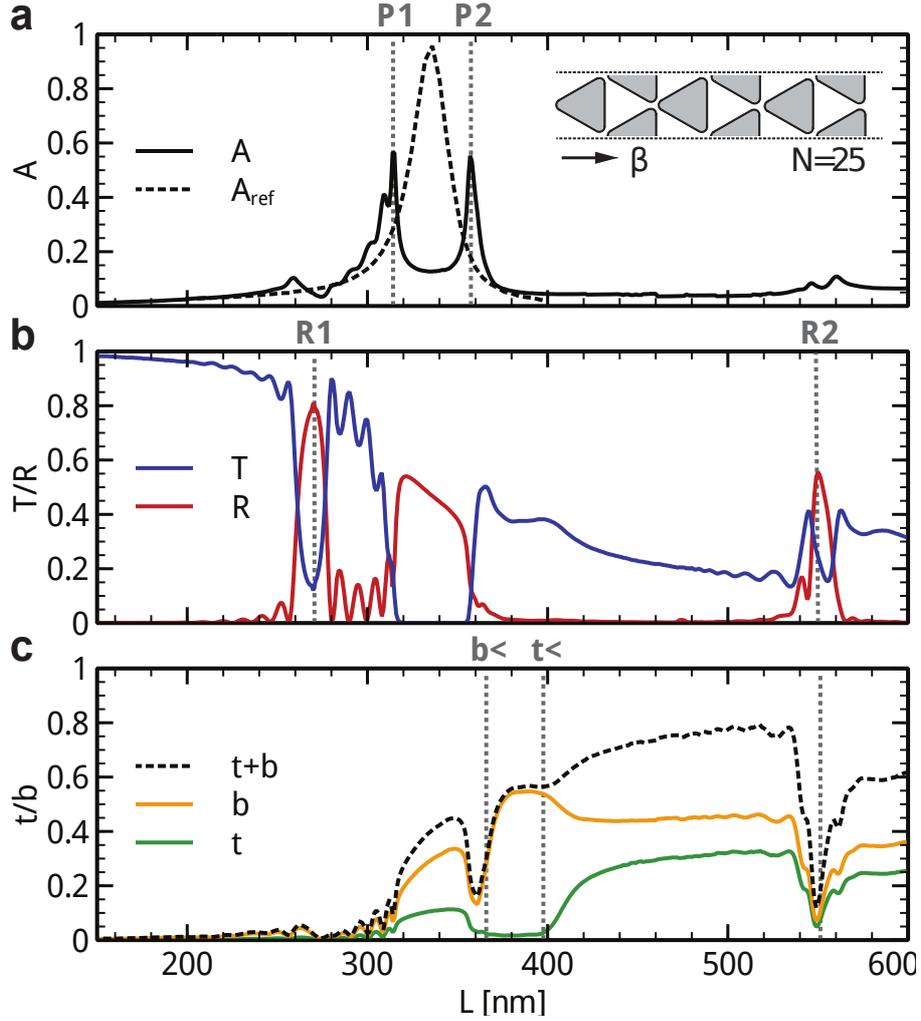}
\par\end{centering}

\caption{\label{fig:5-TilingScatter}Absorption and scattering coefficients
of the triangular NGT recorded over pitch length $L$. (a) Absorption
coefficient A of the NGT for in-plane (solid) and out-of-plane (dashed)
excitation; (b) In-plane transmission T (blue) and and reflection
R (red) coefficients; and (c) coefficients for out-of-plane scattering
into the $\mathrm{SiO}_{2}$ substrate b (yellow) and air superstrate
t (green) and the sum t+b (dashed). The colors of the lines match
those of \ref{fig:3-TrimerScattering}a.}
\end{figure}

\paragraph*{Nano-Gap Tiling: Scattering Spectra.}

Having established how light is focused and scattered by a single
trimer molecule on the surface, we proceed to study the optical characteristics
of triangular nano-gap tiling (NGTs), which can be understood as a
dense array of coupled trimer molecules. These structures are of particular
practical interest as they allow for in-plane excitation of a dense
array of bright spots accessible from the surface. For the following
we chose a laterally periodic array of $N=25$ triangular elements
in propagation direction. To gain an understanding of the photonic
and plasmonic resonances of integrated NGTs, we first compare the
change of absorption of the triangular NGTs for out-of-plane excitation
(\textit{via} total internal reflection) and in-plane excitation (from
the waveguide) with varying element pitch $L$.

\ref{fig:5-TilingScatter}a (solid black line) shows the calculated
absorption spectrum for in-plane excitation for buffer and waveguide
heights of $b=50\,\mathrm{nm}$ and $w=220\mathrm{\, nm}$, respectively.
The absorption spectrum shows two distinct peaks, marking, as can
be seen from the spectra in \ref{fig:5-TilingScatter}b, the edges
of a transmission stop-band. To understand the origin of the hybridization
in the waveguide-integrated configuration, we compare the in-plane
absorption spectrum with the absorption of the NGT under out-of-plane
excitation. We place the NGT on a thick buffer of $\mathrm{SiO_{2}}$
thereby isolating the NGT from the influence of the silicon layer.
To recreate the same conditions of excitation as for in-plane excitation
we inject a plane wave from a semi-infinite $\mathrm{Si}$ substrate
(prism) below the buffer at an incident angle of $\theta_{inc}=\arcsin(n_{eff}/n_{\mathrm{Si}})$.
At this angle the injected plane wave experiences total internal reflection
at the $\mathrm{Si}$/$\mathrm{SiO_{2}}$ interface and can only evanescently
couple to the surface tiling with an in-plane projection of the wave
vector that precisely matches the propagation constant $\beta=n_{eff}k_{0}$
of the waveguide mode. The result, plotted as dashed line in \ref{fig:5-TilingScatter}a,
shows a single broad plasmonic resonance at around $L\approx340\,\mathrm{nm}$
centered between the two in-plane absorption peaks. The splitting
of the plasmonic resonance and the emergence of a transmission stop-band
is a result of the interaction between the plasmonic resonances of
the tiling with the propagating mode in the waveguide.  When the
trimer molecules are coupled in a dense array, the LSPRs couple to
a plasmonic band that supports propagation of SPPs along the chain
of nanoparticles.\cite{Maier2003b} The mutual coupling of the propagating
plasmonic and photonic modes leads to the formation of waveguide-plasmon
polaritons, which, due to the strong coupling (low buffer height),
anti-cross at the point where the frequencies and $k$-vectors of
the plasmonic and waveguide modes match.\cite{Christ2003a} As a consequence
of the mode hybridization, the observed mode-splitting increases with
decreasing buffer size as the coupling parameter between the waveguide
and the surface plasmonic modes become exponentially stronger. 

A fundamental difference between in-plane and out-of-plane excitation
concerns the number of available scattering channels. While plane-wave
ATR spectroscopy measurements record energy transfer into absorption
(A), transmission (T) and reflection (R) channels, one must, for integrated
designs, also consider the scattering into the superstrate (t; top)
and the substrate (b; bottom). In \ref{fig:5-TilingScatter}b/c we
show the scattering coefficients of the triangular NGT over $L$ in
the range $200\,\mathrm{nm}$ to $650\,\mathrm{nm}$. With the structure
becoming resonant to the effective wavelength ($L\gtrsim\lambda_{eff}/2$),
it acts as a plasmonic crystal.\cite{Barnes1996,Salomon2001,Zayats2005}
Both plasmonic and photonic scattering are mediated by the triangular
lattice of the NGT. In the propagation direction, the projection of
the incident wave vector (the propagation constant $\beta$) on the
reciprocal lattice vectors ($\vec{a}_{1}$ and $\vec{a}_{2}$ ) are
equal. The $k$-matching condition $k_{s}=\beta\pm m2\pi/L$ dictates
that the in-plane projection of the scattered wave vector equals the
incident wave vector $\beta$ plus multiples $m$ of the effective
lattice vector $2\pi/L$. Using $k_{s}=n_{s}k_{0}\sin(\theta_{s})$
we associate $k_{s}$ with the scattering angle $\theta_{s}$ relative
to the surface-normal and the refractive index $n_{s}$ of the scattering
channel. Resolving this condition with respect to $L$ gives
\begin{equation}
L=\pm m\frac{\lambda_{0}}{n_{eff}-n_{s}\sin(\theta_{s})}\label{eq:kmatching}
\end{equation}

Apart from the plasmonic stop-band between $L_{P1}$ and $L_{P2}$
(\ref{fig:5-TilingScatter}a), there are two more stop-gaps (at $L_{R1}$
and $L_{R2}$) in the spectrum (see \ref{fig:5-TilingScatter}b),
which can be explained by resonant back-scattering. Setting $\theta_{s}=-\pi/2$
and $n_{s}=n_{eff}$ in (\ref{eq:kmatching}) gives $L=m\lambda_{0}/(2n_{eff})$,
the Bragg conditions for the first and second ($m=1,\,2$) order reflection
peaks at $L_{R1}\approx280\,\mathrm{nm}$ and $L_{R2}\approx559\,\mathrm{nm}$,
which are in excellent agreement with \ref{fig:5-TilingScatter}b.
In contrast, out-of-plane scattering into the $\mathrm{SiO_{2}}$
substrate ($n_{s}=1.44$) and air ($n_{s}=1.0$) can occur over the
whole lower and upper half-space. Again using equation (\ref{eq:kmatching}),
we mark the points for $\theta_{s}=-\pi/2$ as $L_{t<}$ and $L_{b<}$
for scattering to the top and bottom respectively. We also note that
the second Bragg peak at $L_{R2}\approx559\,\mathrm{nm}$ always coincides
with normal-to-plane scattering into the substrate and superstrate,
reducing the proportion of energy transmitted into these channels.

\begin{figure}[H]
\begin{centering}
\includegraphics[width=0.75\textwidth]{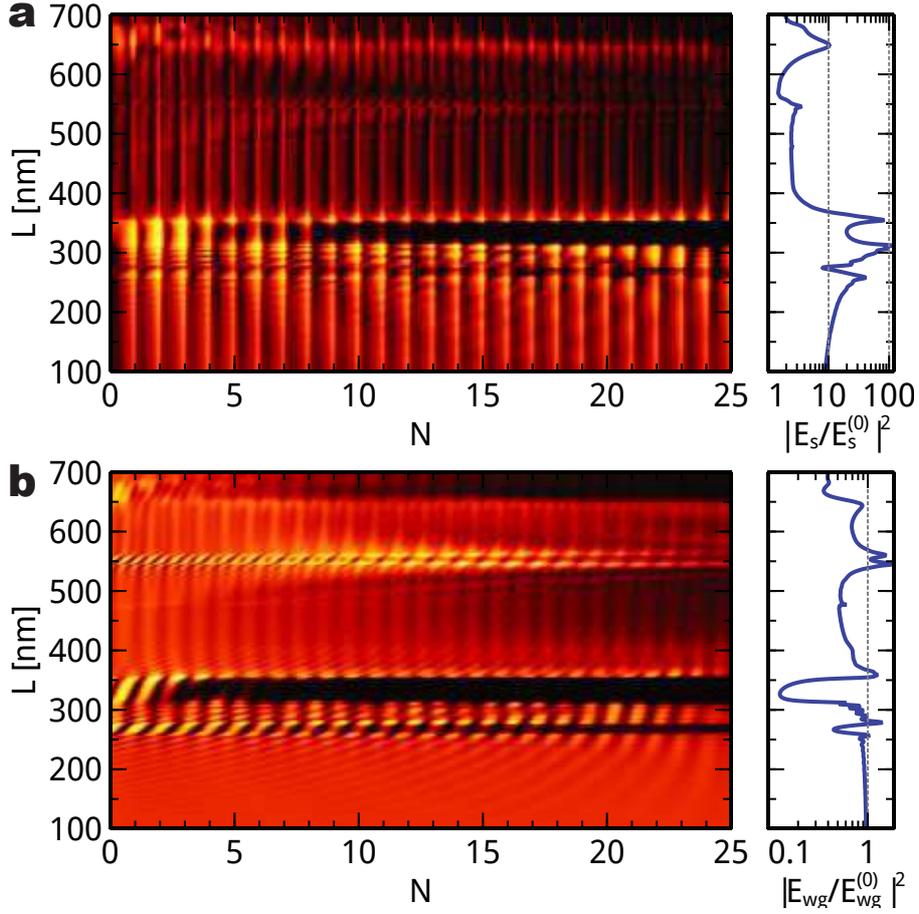}
\par\end{centering}

\caption{\label{fig:6-TilingEnhance}Log plots of the laterally (along $x$-direction)
averaged field intensities (a) the surface ($y=y_{s}$) and (b) in
the center of the waveguide ($y=y_{wg}$) for a NGT structure with
$N=25$ elements. The corresponding longitudinally averaged intensities
are plotted on the right hand panels.}
\end{figure}

\paragraph*{Nano-Gap Tiling: Surface Field Enhancement.}

Changing the pitch $L$ of the NGT does not only affect the scattering
of energy into the various channels but has a strong impact on the
concentration of energy on the surface as the plasmonic resonance
and localization of fields in the waveguide changes. To investigate
the enhancement of fields we (as before) inject a waveguide mode into
a NGT with $N=25$ elements and record the laterally (along $x$-direction)
averaged intensities on the surface $|E_{s}(z)|^{2}=|E(y_{s},z)|^{2}$
and in the center of the waveguide $|E_{wg}(z)|^{2}=|E(y_{wg},z)|^{2}$.
To obtain the relative field intensities on the surface and inside
the waveguide, we divide by $|E_{s}^{(0)}|^{2}=|E^{(0)}(y_{s})|^{2}$
and $|E_{wg}^{(0)}|^{2}=|E^{(0)}(y_{wg})|^{2}$, the bare intensities
without the NGT on the surface (\ref{fig:6-TilingEnhance}; left panels).
Finally, to quantify the net enhancement of the field we average the
normalized intensities on the surface and inside the waveguide along
the propagation direction (\ref{fig:6-TilingEnhance}; right panels).
The results show that the normalized surface intensities peak at
the hybrid resonances at $L_{P1}$ and $L_{P2}$. Both the plasmonic
stop gap and the Bragg resonances inhibit propagation in the waveguide,
leading to strong attenuation of the fields in the propagation direction.
However, owing to the excitation of SPPs, energy keeps propagating
in form of SPPs on the surface with less attenuation, leading to an
elevated surface field enhancement in these spectral regimes (compare
\ref{fig:6-TilingEnhance}a and \ref{fig:6-TilingEnhance}b).

For pitch lengths above the plasmonic resonances the surface intensities
weaken due to the onset of out-of-plane scattering. In contrast, below
the first Bragg resonance, in the subwavelength regime, the strength
of the surface intensities settles at an average value of $\sim10$,
while the average intensity in the waveguide remains $\sim1$ across
the length of the NGT. This off-resonant enhancement of surface intensities
can be attributed to the lightning rod effect,\cite{Gersten1980}
which leads to a concentration of field strength into the focal points
of the NGT without inducing high plasmonic losses.

For functionalization of the surface (\textit{e.g.} with polymers,
dyes, \textit{etc.}) we are interested in the effective enhancements
of the linear and nonlinear susceptibilities of the material deposited.
To find the enhancement for a particular process, we divide the $n$'th
power of the normalized surface field by the normalized electric field
within the waveguide and integrate over the area of the NGT. For example
the effective intensity enhancement ($n=2$) is given by

\begin{equation}
\eta_{eff}^{(I)}=\frac{1}{A}\int_{A}\frac{|E(x,y_{s},z)|^{2}}{|E(x,y_{wg},z)|^{2}}dxdz
\end{equation}

\begin{figure}[H]
\begin{centering}
\includegraphics[width=0.75\textwidth]{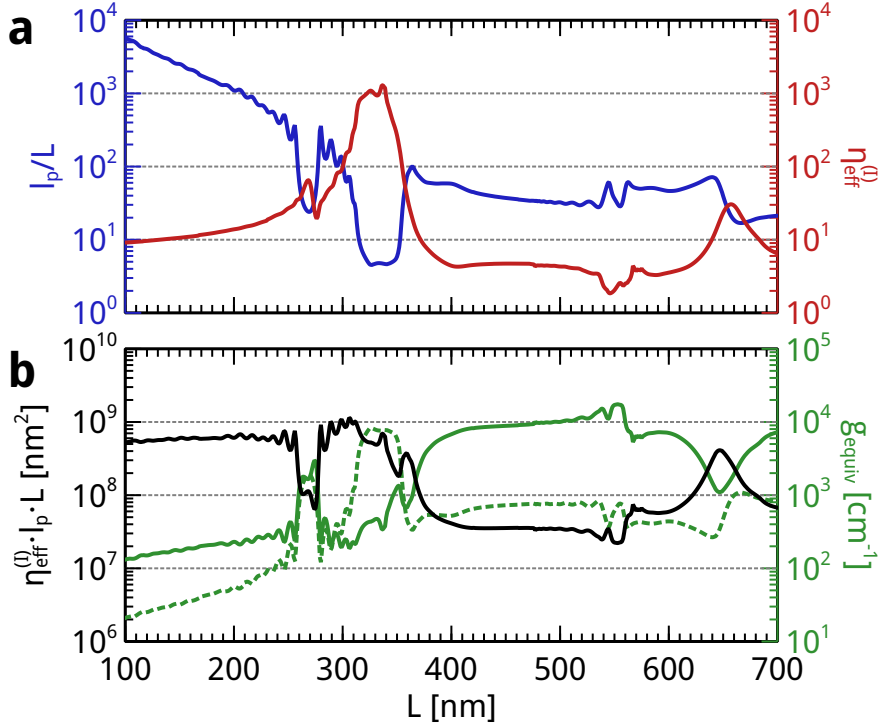}
\par\end{centering}

\caption{\label{fig:7-TilingFOM}(a) The propagation length of the TE mode
in the waveguide (blue line) relative to the element pitch and the
effective surface intensity enhancement (red line). (b) The effective
attenuation cross-section $\sigma_{eff}=\eta_{eff}^{(I)}l_{p}L$ (black
line) and the equivalent gain coefficients (green lines) required
for full loss-compensation. The results for gain on the surface (solid
green line) and in the waveguide (dashed green line) differ due to
the impact of confinement factor and field enhancement.}
\end{figure}

The fact that we divide the surface intensities at every point by
the intensity in the waveguide underneath allows us to eliminate propagation
effects such as attenuation, back-reflection and standing wave patterns.
This approach is valid as the evanescent interaction of the waveguide
fields with the surface structure does not involve significant retardation
effects. The so-obtained effective enhancement of the surface intensity
is plotted in \ref{fig:7-TilingFOM}a (red line) together with the
unit-less effective propagation length $l_{p}/L$ (blue line) over
the pitch $L$. One notices that, in the off-resonant regime, the
per element propagation length decreases exponentially for decreasing
pitch values, while the effective enhancement of the surface intensity
asymptotically approaches a value of $\sim10$. When the pitch approaches
the plasmonic resonance of the NGT at around $L\approx340\,\mathrm{nm}$,
the effective propagation length is at a minimum as the plasmonic
stop gap prevents propagation of the wave. The effective intensity
enhancement is strongest in this regime as the NGT is excited resonantly,
leading to high field concentrations (relative to the waveguide fields)
in the gaps that propagate in form of SPPs along the surface.  For
pitch values larger than $L\approx400\,\mathrm{nm}$, where out-of-plane
scattering dominates, the values of $\eta_{eff}^{(I)}$ and $l_{p}/L$
change only slightly with the pitch. 

\paragraph*{Nano-gap Tiling: Equivalent Gain.}

Although the achieved intensity enhancement in the subwavelength regime
is comparably low, the propagating photonic field experiences low
attenuation and remains in interaction with the NGT. To quantify the
trade-off between field concentration and attenuation we introduce
an effective attenuation cross-section
\begin{equation}
\sigma_{eff}=\eta_{eff}^{(I)}l_{p}L
\end{equation}

which is shown in \ref{fig:7-TilingFOM}b (black line). Strikingly,
this cross-section is close-to-constant in the subwavelength regime
(below $L=250\,\mathrm{nm}$) and in the radiative regime before the
second Bragg resonance ($L=400-550\,\mathrm{nm}$). In these regimes
\begin{equation}
\eta_{eff}^{(I)}=\frac{const}{L^{2}}\cdot\left(\frac{l_{p}}{L}\right)^{-1}
\end{equation}

which means that the ability of the NGT to concentrate energy on the
surface is proportional to the density of hot-spots ($\sim1/L^{2}$)
and the normalized attenuation ($L/l_{p}$). This relation is of particular
relevance in the context of device functionalization with nonlinear\cite{Baumberg2005}
and gain materials.\cite{Wuestner2010,Hamm2011,Pusch2012} As an application,
we imagine an active material deposited on the surface. To fully compensate
the dissipative and radiative losses we require an equivalent gain
of
\begin{equation}
g_{equiv}=\frac{1}{\Gamma\eta_{eff}^{(I)}l_{p}}
\end{equation}

introducing the confinement factor
\begin{equation}
\Gamma=\frac{\int_{-\infty}^{+\infty}\mathrm{d}y\Theta(y)|E^{(0)}(y)|^{2}}{\int_{-\infty}^{+\infty}\mathrm{d}y|E^{(0)}(y)|^{2}}
\end{equation}

which quantifies the proportion of energy that interacts with the
gain region ($\Theta(y)=1$ in the gain-filled section and 0 elsewhere).
For the given structure we find that $\Gamma\approx0.0143$ for a
layer of $30\,\mathrm{nm}$ gain media deposited on the buffer, in
between the elements of the plasmonic NGT. The retrieved equivalent
gain values are plotted as green line in \ref{fig:7-TilingFOM}b.
In the subwavelength regime where the attenuation cross-section reaches
a value of $\sim4-5\times10^{8}\,\mathrm{nm}^{2}$ there is a linear
increase of the equivalent gain with $L$ from $g_{equiv}\approx150\,\mathrm{cm}^{-1}$
at $L=100\,\mathrm{nm}$ to $g_{equiv}\approx300\,\mathrm{cm}^{-1}$
at $L=250\,\mathrm{nm}$. At the first Bragg resonance ($L\approx270\,\mathrm{nm}$)
the gain values peak at $g_{equiv}\approx2000\,\mathrm{cm}^{-1}$
before they drop again to values of only $g_{equiv}\approx200\,\mathrm{cm}^{-1}$
in the plasmonic stop gap, where the propagation constant is lowest
but the effective enhancement highest (see \ref{fig:7-TilingFOM}a).
Beyond the plasmonic resonances, radiative loss dominates and the
field enhancement is low. This leads to high equivalent gain values
of $\sim10^{4}\mathrm{cm}^{-1}$ that would be required to compensate
attenuation losses. The results illustrate that loss-compensation
in the proposed hybrid structure is a realistic prospect even in the
resonant plasmonic regime. Finally, we note that using epitaxially
grown InGaAsP based semiconductors structures gain can also be introduced
directly into the waveguide structure. In this case (keeping the geometric
dimensions) the confinement factor is $\Gamma\approx0.82$ but there
is no plasmonic field enhancement, resulting in the dashed green line
in \ref{fig:7-TilingFOM}b. While the required gain for loss-compensation
is generally lower, this is particularly not true in the regime where
the waveguide mode couples to the surface plasmon resonances of the
surface.

\subsubsection*{CONCLUSION}

In conclusion we reported on a new class of low-loss integrated surface
structures, plasmonic nano-gap tilings (NGTs), which can concentrate
light on the surface of a photonic chip in a lattice of bright-spots.
To understand their unique properties, we investigated triangular
trimers in an integrated configuration, showing their potential to
scatter and concentrate light on the surface of the chip in dependence
of their size. For the assembled NGT we identified different regimes
of operation depending on the pitch length: the off-resonant regime,
where the NGT acts as true light-concentrating metasurface, and regimes
dominated by in-plane (back) scattering, excitation of plasmonic modes,
and out-coupling of radiation. Using a novel technique for the extraction
of the field enhancement, we were able to define an effective attenuation
cross-section, which plateaus in the off-resonant and radiative regime.
The equivalent gain required for loss-compensation on the surface
and in the waveguide illustrate that loss-compensation in these structures
is realistically possible. The triangular structure that forms the
basis of this work is one of many possibilities to cover a surface
with a regular 2D tiling. The chosen combination of particle shape
(triangles) and crystal lattice (hexagonal) was guided by our aim
to concentrate light in a dense array of hot-spots on the surface.
Tilings with higher symmetry, such as a checkerboard tiling (diamonds
on a square lattice), are possible but do not achieve the same surface
field enhancement, as, for a given propagation direction, only half
of the nano-gaps are excited. We note that waveguide-integrated plasmonic
surface arrays are not limited to the passive SOI platform, but can
be also be integrated on active semiconductor material systems (\textit{e.g.} 
III-V semiconductors). While surface functionalization with gain
materials were considered in this work, other nonlinear materials
can be conveniently deployed on the surface. The observed hybrid waveguide
plasmon-polariton resonances are also highly sensitive to changes
in the dielectric constant\cite{Zentgraf2009} and could be employed
for waveguide-integrated dielectric sensing applications that would
additionally benefit from the field enhancement on the surface. In
combining the low-loss, light-enhancing and resonant properties of
plasmonic and photonic structures with NGT structures may offer new
routes to hybrid devices, such as on-chip sensors, plasmonic couplers,
active surface emitters, switches and all-optical modulators.

\subsubsection*{METHODS}

\paragraph*{Numerical Simulations.}

All numerical simulations were performed in frequency domain using
the full-vectorial finite-element method. To calculate the spectras
of the isolated trimer elements we carried out scattering simulations
with COMSOL using a two-step process: First we solved for the background
field of the waveguide without the antenna using a combination of
port boundary conditions (at the front and back) for waveguide mode
injection, perfect electric boundary conditions (on the left and right)
and perfectly matched layers (PMLs) on top and bottom. The calculation
of the background field was followed by scattered-field calculations
where the antenna was placed onto the waveguide. For this second step
the PEC and PML boundaries have been replaced by PMLs enabling outward
propagation of the scattered fields in all directions to determine
the scattered energy flux. For the simulation of the finite NGT, we
performed single step simulations using JCMwave. Periodic boundary
conditions (PBC) were used for the sides of the structure to simulate
a laterally infinite tiling. PMLs were used for the continuation of
the sub- and superstrate layers, and the continuation of the forward
and backward waveguide stack . The distance of the PMLs was chosen
sufficiently large to avoid spurious backreflections and to accurately
resolve the out-of-plane energy flow and the in-plane transmission
and reflection coefficients.

 The simulations were carried out for perfectly periodic structures
and for a fixed direction of excitation by injection of a TE waveguide
mode along the $z$-axis. To investigate off-axis propagation and
the impact of imperfections, such as spatially varying gap sizes or
corner radii, one would need to conduct systematic studies on laterally
and longitudinally large (and finite) structure samples, which are
subject to future works. 

\textit{Acknowledgement:} The authors would like to thank the EPSRC
and the Leverhulme Trust for supporting this work.

\providecommand*{\mcitethebibliography}{\thebibliography}
\csname @ifundefined\endcsname{endmcitethebibliography}
{\let\endmcitethebibliography\endthebibliography}{}

\end{document}